\documentclass[aps,pre,groupedaddress,amsmath,showpacs, twocolumn]{revtex4} 

\bibpunct{}{}{,}{s}{}{}

\usepackage{dcolumn} \usepackage{graphicx} \usepackage{natbib}

\usepackage{color}

\newcommand{\ecsend}[1]{#1}

\newcommand{\besm}[1]{\textbf{#1}}

\begin{document}

\title{Brief, embedded, spontaneous metacognitive talk indicates thinking like a physicist} 
\date{\today}

\author{Eleanor C. Sayre} 
\affiliation{Department of Physics, Kansas State University, 116 Cardwell Hall, Manhattan, Kansas 66506-2601}

\author{Paul W. Irving} 
\affiliation{Department of Physics and Astronomy, Michigan State University, 567 Wilson Road, East Lansing, Michigan 48824}


\begin{abstract} 
Instructors and researchers think ``thinking like a physicist'' is important for students' professional development. However, precise definitions and observational markers remain elusive.  We reinterpret popular beliefs inventories in physics to indicate what physicists think ``thinking like a physicist'' entails.  Through discourse analysis of  upper-division students' speech in natural settings, we show that students may appropriate or resist these elements.  We identify a new element in the physicist speech genre: brief, embedded, spontaneous metacognitive talk (BESM talk). BESM talk communicates students' in-the-moment enacted expectations about physics as a technical field and a cultural endeavor.  Students use BESM talk to position themselves as physicists or non-physicists.  Students also use BESM talk to communicate their expectations in four ways: understanding, confusion, spotting inconsistencies, and generalized expectations.
\end{abstract}


\maketitle


\section{Julie}
\label{sec:Julie}
Consider the following case of ``Julie''\footnote{Names are pseudonyms.}, a student in an intermediate Classical Mechanics course at a large research institution in the United States.  Julie is a conscientious student who frequently works on her homework with her peers and a teaching assistant.

In class, Julie's professor solved a linear first order differential equation which has exponential solutions.  He remarked that some solutions ``blow up'' for large values, which confused Julie.  She recounts the interaction to the TA, whose tone is friendly in this interaction:
\begin{description}
\item[Julie:] [The professor] said ``blows up" and I was thinking, when I think blow up I think like you know (explosion sound effect)\dots [it's] funny to be like what the heck and then ask him and he's like oh I mean this and I'm like oh okay
\item[TA:]\dots pretty much everybody in physics uses ``blows up" to mean the exponential gets really big.
\item[Julie:] well\dots \besm{I don't know enough physics apparently.}
\end{description}

Julie's last statement is wry and dejected.  It is a generalized assessment of her knowledge of physics which links both the technical aspects of learning physics (here, solutions to a differential equation) with the social aspects (here, knowing the physics idiom ``blows up'').  Her statement is spontaneous: no one asked her to recount this story, nor to reflect on her understanding of physics. It is embedded in a larger discussion about solving differential equations. Far from a long soliloquy, these six words suggest a lot about how Julie views herself in relation to the practice of physics, and it has disappointing implications for her persistence in the field. 

In this paper, we argue that the type of talk exemplified by this short exchange -- brief, embedded, spontaneous metacognitive talk -- is important to many educationally significant settings across science fields and throughout science curricula. It also indicates cultural aspects of belonging to a field and becoming a physicist. 

\section{Overview}
Becoming a physicist is an arduous process in which students study physics content, grow in mathematical sophistication, and develop their research interests and expertise. These technical goals accompany social and cultural ones: learning how to communicate with other physicists, joining physics communities of practice, and developing an identity as a physicist.  In short, to become physicists, students need to think like physicists.  

Both instructors and researchers believe that ``thinking like a physicist'' (TLP) is important for students' professional development; however, precise definitions and observational markers remain elusive.  The diverse descriptions of TLP share a common bond: to think like a physicist, a student needs to share the epistemology of a physicist. We look at epistemological beliefs and commitments instead of content knowledge for three reasons: because they indicate a deeper level of understanding physics than mere conceptual understanding; because they are important for student enculturation and persistence; and because they make comparisons fruitful between upper-level students and professionals. 

In this paper, we critically review literature from PER and cognate fields to schematize the epistemological beliefs and commitments that make up TLP. 

We are particularly interested in how students' spontaneous discourse in natural settings reflects their epistemological beliefs and commitments.  Towards that goal, we analyze observational data of upper-level students in laboratory and theory classes, and during homework discussion periods.  We conduct discourse analysis to find their enacted epistemological commitments.  What do these students think that doing physics entails? How do they know when they are on the right track?  In contrast to research at the introductory level on students' professed beliefs (such as with beliefs surveys or in interviews), we are interested in how students' epistemological commitments are evident ``in the wild.''

We offer brief, embedded, spontaneous metacognitive (``BESM'') talk as one indicator of students' enacted epistemological commitments, and compare their epistemological commitments to those of professional physicists through critical review of the literature. Our primary goal in this paper is to raise the specter of measuring TLP in students' in-the-moment discourse.   We motivate our discussion of BESM talk through theory and exemplars. We do not attempt an exhaustive catalog of phrases that ``count'' as BESM talk, nor do we measure the prevalence of it in natural speech across instructional settings.    

\section{Schematizing ``Thinking Like a Physicist''}
Thinking like a physicist (TLP) is deeply important to educators in physics: it is often an explicit learning goal, or forms an integral part of the hidden curriculum\cite{Beichner2007, Crouch2010, Chasteen2012}.  However, as a construct, it eludes definition, and we often allude to it obliquely.  In this section, we attack this problem in two wholly separate ways: an examination of the literature around TLP, and a reinterpretation of popular beliefs surveys in physics.

\subsection{Literature}
 
In an early landmark paper, Van Heuvelen\cite{VanHeuvelen1991} described TLP as a way of solving problems in which physicists first construct qualitative relationships of physical processes and problems, reason about those relationships and mathematize them, then solve the problem.  He argues that students often skip the beginning parts of this process because they are never given the opportunity to practice it, and argues that they are integral to TLP. Other accounts of TLP describe it as a set of skills, abilities, traits, or attitudes and different types of thinking \cite{St.John1980, Cerny2007, VanZee2010, Grayson2006, Sabella2010, Demaree2010, Chasteen2012, Redish2012}. 
 
TLP is also generally considered a learning goal in the hidden curriculum (especially in laboratory learning environments and the upper division) \cite{Ivany1968, Beatty2002, Beichner2007, Crouch2010, Chasteen2012}, though occasionally it is made explicit\cite{VanZee2010}.  As a learning goal, TLP is  related to moving students from aspiring physicists to professional physicists \cite{Karagiorgi2005, Cerny2007}. 

For all that TLP is important to physicists and physics educators, it is often understood and encouraged by analogy.  Airey and Linder\cite{Airey2009} liken learning physics to learning a foreign language.  Students need to pick up a multimodal discourse around physics, where each mode is a particular way of apprehending the content: verbally, mathematically, diagrammatically, with equipment, etc.  By analogy to second-language learning, they contend that the most efficacious way to learn physics is by immersion in genuine practices with the natives, such as by participating in research with professional physicists.  In their framework, to TLP is to adopt the discourse of physicists across a ``critical constellation of modes'', and a necessary prerequisite to becoming a physicist is to be fluent in that discourse. 

In contrast, Van Heuvelen's focus on improving students' understanding of physics causes him to liken learning physics to impedance matching between instruction and students.  In this metaphor, formal education is a transformer. To deliver maximal power (information) to the load (student), the transformer must be ``attuned to the characteristics of student minds at all times.''\cite{VanHeuvelen1991}  This metaphor masterfully elicits cultural ways of knowing physics, exploiting TLP to explain how to teach readers about TLP.  

Within the physics education literature and the minds of many physicists, the definition of TLP maintains an element of ambiguity in the general case,\cite{Redish2013a,Hammer1996} even after 25 years of work. In the words of the United States Supreme Court, we ``know it when [we] see it''\cite{Jacobellis1964}. 
However, the diverse descriptions of TLP share a common bond: to think like a physicist, a student needs to share the epistemology of a physicist.  

Researchers in the philosophy and sociology of physics have studied physicists' enacted beliefs and commitments in research labs (e.g. \cite{Traweek1988} and in physicists' discourse (e.g. \cite{Mitroff1974}).  Nevertheless, ``what it means to `think like a physicist' is not nearly as well developed as our understanding of physical phenomena''\cite{Hammer1996}, especially among the researchers and faculty who interact with physics classrooms where undergraduates tend to gather.   Physicists' epistemological beliefs and commitments vary by setting (just like all other humans), so we focus our attentions on how TLP may be expressed for the classroom.  

We study epistemological commitments instead of content knowledge for all the reasons that people want to study students' beliefs: because they are important for retention and persistence, because they indicate a deeper level of what ``getting it" means than mere conceptual understanding, and because we can study them across populations with wildly divergent content knowledge. 

\subsection{Beliefs Inventories}

We reinterpret two popular epistemological beliefs inventories through the lens of schematizing what it means to think like a physicist in undergraduate coursework. 

The Colorado Learning Attitudes about Science Survey (CLASS)\cite{Adams2006} and the Maryland Physics Expectations Survey (MPEX)\cite{Redish1998} are two of the most widely-used beliefs surveys in physics\cite{Madsen2014CLASS}.  Their popularity is one of their major strengths among faculty who want to compare their students to others'.

In these surveys, students respond to Likert-scale statements about their experiences in physics classes or with physics as a discipline, their beliefs about physics, and their beliefs about success in classes.  Responses are scored based on how well students agree with physics experts. ``Favorable'' responses mean that a student professes a more physicist-like belief; ``unfavorable'' ones the opposite belief.  

Both of these instruments were validated through interviews with and administration to physics faculty and related experts in teaching physics to undergraduates, as well as other methods.  A full discussion of their validation strategies is beyond the scope of this paper. However, we note that the instruments themselves represent curated and expert-validated lists of statements about what it means to think like a physicist in an undergraduate classroom context. 

From these beliefs surveys, we notice that epistemological commitments in undergraduate coursework associated with TLP include similar dimensions.  These dimensions are not wholly distinct, and individual items on the CLASS test several of them at once.  

\begin{description}
\item[Physics in the real world:] Physicists generally believe that the material in physics classes is generally applicable outside of them as well.
\item[Connections among physics:] Physicists generally believe that a small number of physical principles are applicable in a wide variety of scenarios, and that physics topics are connected to each other in deep and meaningful ways. 
\item[Sense making:] Physicists generally believe that it is important to make sense of physical scenarios, as opposed to blindly applying formulas and procedures.  Sense making is sometimes conceptualized as expansive framing\cite{Irving2013b}, or as an effort to coordinate multiple modes or representations\cite{Airey2009, VanHeuvelen1991, Gire2013a}, or as constituted in part by mechanistic reasoning\cite{Russ2006}.
\item[Problem solving:] Physicists generally believe that solving physics problems, particularly those assigned in classes, is an important part of TLP at the undergraduate level. Often, but not always, the appropriate and mindful use of mathematics is part of problem solving.
\item[Effort and aptitude:] Physicists generally believe that anyone can do physics if they apply themselves and work hard.  In contrast, some students believe that only some special people can do physics, or that physics ability is an inborn trait rather than a developable skill.
\end{description}

From these categories, we notice that physicists believe that physics class material is both well-connected to itself and to non-class material through broadly applicable principles, and that to understand physics involves effortful, mindful problem solving.  Physicists emphatically do not endorse a fixed mindset\cite{Dweck2006Mindset}, rote formula or procedure use, or compartmentalization of physics classroom material.  

We interpret the dimensions on these surveys as a kind of ``basis set'' for what TLP means in an undergraduate classroom context.  Other dimensions in this set may be possible, but given the validation work for these surveys, their agreement with each other, and their popularity, we suspect that these categories are the most prevalent in and important to our community.
Section \ref{sec:besmkinds} relates these categories to the kinds of BESM talk.

\subsection{Inventories or natural settings?}

However, as a method for acquiring actionable, timely information from students, surveys have some natural drawbacks.  They are typically administered at the start and end of each semester, and students' responses are examined in the aggregate.  For classroom educators who want to support students' epistemologies in the moment, surveys are too far removed and aggregated.  BESM talk, by virtue of being embedded in natural settings, is an alternative measure that is accessible in the moment of teaching. Additionally, because it is spontaneous, a researcher need not pull students out of of their natural settings in order to study it: classroom video of student discourse suffices.  

The surveys have another substantial methodological problem for measuring beliefs and behavior ``in the wild'': they measure only students' \textit{professed} beliefs.  Students sometimes exhibit differences in what they ``really believe'' (as compared to what they think a physicist would say)\cite{Gray2008a}. While some research suggests that these ``splits'' are minimal\cite{Adams2006}, other surveys have chosen to embrace the difference\cite{zwickl:442}.  In either case, students' in-the-moment, enacted beliefs may differ from their professed beliefs.  For example, a student may respond favorably to ``Knowledge in physics consists of many disconnected topics'' on the CLASS, but in practice treat the material in each chapter as if it is unrelated to material in nearby chapters.   Surveys, as a methodology, cannot capture enacted beliefs, only professed ones. 

In this work, we are concerned with students' enacted beliefs as they perform physics in natural settings.  Because of these two natural limitations on beliefs surveys, they are ill-matched for our work (even as these particular surveys are excellent for their intended purposes).  

\section{Talking like a physicist}
An accessible marker of \textit{thinking} like a physicist is \textit{talking} like a physicist\cite{Manogue2010, VanZee2010}.  By examining physics students' speech, we uncover their physics expectations and how they position themselves as regards the study of physics and physics as a prospective professional community.\cite{Lave1991, Wenger1998}  

Of course, students might not talk like physicists: they may position themselves in opposition to physics culture, or their ideas of what constitutes doing ``good physics'' may not align with what physicists think.  These differences interest us because they help paint a richer picture of students' enculturation into physics as a technical field and community endeavor.

\section{Metacognitive talk}
As a learning goal related to the development of professional physicists, TLP helps develop students' metacognition\cite{Manogue2010}.  Metacognition is often described as ``second-order'' cognition\cite{Weinert1987, Butterfield1995}: thoughts about thoughts, knowledge about knowledge, or reflections upon one's actions\cite{Tsai1998}. Like TLP, it is often perceived as an explicit learning goal or is a part of the hidden curriculum.  Unlike TLP, metacognition is important to a broad range of degree programs and learning environments. Metacognitive development is typically seen as a necessary step in becoming a life-long learner. 

Previous research on metacognitive talk in math and physics has focused on teaching students to monitor their reasoning or self-regulate their learning (e.g. \cite{Koch2000, Zion2005,Georghiades2004}). Generally, increased metacognition results in increased success in a given instructional context\cite{Tsai2001}, though some research indicates that it is not the \textit{quantity} of metacognition that matters, but whether it changes students' behavior\cite{LippmannKung2007a}.  In this paper, we examine a particular kind of metacognitive talk which is common across many learning environments in which students do physics.  It is rare in research on metacognition to focus on metacognitive talk in natural settings (notable exceptions include \cite{LippmannKung2007a, Winne1996, Georghiades2004}), but we are especially interested in how students actually do science away from education research laboratories.  

We examine both the physics being discussed and the metacognitive talk intertwined with it. Unlike research on self-regulation, our research posits that metacognitive talk can indicate how students are taking up the epistemological commitments of physicists,\cite{Tsai2001} and therefore to what extent they are becoming physicists.  While previous research has focused on categorization at the episode level\cite{LippmannKung2007a} and students' framing\cite{Tannen1987, Hutchison2009, Hammer2004, Berland2012}, we focus on phrases in students' discourse as indicative of their metacognition.

Our research occurs in natural settings for physics students: in classrooms and instructional laboratories, and while they do their homework.  Their metacognitive talk is necessarily \textit{embedded} in these contexts.  Because there is no researcher interaction with the students -- and little instructor interaction -- their talk is \textit{spontaneous}.  In these natural settings, students do not often embark on long, reflective monologues about the nature of physics and their role within it; the talk consists of \textit{brief} phrases interspersed around technical work. Altogether, the kind of talk we investigate is brief, embedded, spontaneous metacognitive talk (``BESM talk'').  This kind of talk has been previously described as ``natural-in-action metacognition''\cite{LippmannKung2007a} in efforts to categorize and quantify the kinds and stages of problem solving in which students engage\cite{LippmannKung2007a, Schoenfeld1992, Artzt1992}.  Our interests are in how students' metacognitive talk reflects (does not reflect) their enculturation into science, specifically physics; physicist-like problem-solving is but one aspect.  

While we believe that students of many disciplines and levels may engage in BESM talk,  our larger research interests are students' professional identity development as physicists.  For this reason, we focus our investigations on journeymen students\cite{Bing2012}: neither introductory students nor professionals, these students' professional identities and ideas about what constitutes good physics are changing rapidly.  A natural consequence of studying BESM talk in this population is that our subject pool is small and our data well-suited to microgenetic learning analysis of video\cite{Parnafes2013, Berland2012, Scherr2009}, a qualitative method related to microgenetic analysis\cite{Siegler2006} which connects cases to theory through close analysis of short video-based episodes. 

Given these constraints, we refine the central research question for this paper: in what ways does students' BESM talk indicate how they are TLP?  In the following sections, we develop an answer through several examples and connections to theory.  We close with a discussion of the practical merits and limitations of this work.  

\section{Speech Genres and physicist language}
\label{sec:theory}

The way we talk -- thematic content, style, compositional structure -- is influenced by the broader context of our speech, including the culture in which we speak.\cite{Bakhtin1986}  The speech and its broader context are often called the ``sphere of communication''\cite{Wertsch1998} or ``social language''\cite{Koschmann1999a}).

Within each social language, relatively stable types of utterances develop, known as speech genres. BESM talk is one such type -- reflective utterances -- embedded in the social language of the physicist.  Speech genres are often considered to be extremely heterogeneous\cite{Bakhtin1986}; however, for the purposes of this analysis we extend the grain size of speech genre as a theoretical construct.  This extension is not unusual in studies based within particular cultures\cite{Roschelle1996}. 

Speech genres differ from Gee's big-D Discourse\cite{Gee2005} in several important ways.  While discourse includes physical artifacts and how they are used as well as language-in-use, speech genres are specialized to both the specific genre and to communicative talk and gesture.  Even small-d discourse, such as championed by Airey and Linder\cite{Airey2009} contains within it non-utterance, non-verbal, non-gestural modes such as use of diagrams or equipment.  In this paper, we focus on the linguistic aspects of students' discourse, and therefore find the idea of speech genres more fruitful. 

Within this framework, students develop as physicists by appropriating the physicist speech genre.  They may start by imitating phrases and discourse through SWOSing\cite{Harlow2006} (Airey and Linder's ``discourse imitation''), but through increased use across multiple modalities and physics contexts, they both master the content and appropriate the speech genre, becoming ``native speakers'' of physics.  

Because students' primary access to the physicist speech genre is through communication with physicists, particularly their professors and near-peers, this work fits neatly into a Communities of Practice\cite{Lave1991,Wenger1998} perspective, where central participants share a speech genre, and peripheral participants must learn the shibboleths in order to become more central.  We ``see'' learning as experience in disciplinary ways of knowing, gained through participation.\cite{Northedge2002} 

At this point, we need to point out a natural limitation on this work.  We are looking at the BESM talk speech genre within the social language of the physicist. If we were examining the talk of chemistry students, we would be investigating the BESM talk speech genre within the social language of the chemist. We would expect the details of the genre to change, though the existence of the genre (and some of the elements) would remain the same. We believe that the construct of BESM talk is applicable cross-culturally, but that changing the context and culture of interest will necessarily change the details of the speech genre.

\section{Methods}

The research presented in this paper is part of an ongoing ethnographic research project attempting to understand how upper-level physics students move along a developmental path toward or away from becoming a physicist.\cite{Irving2014,Irving2104} 

\subsection{Context}
Our research participants are generally physics majors or minors enrolled in middle- and upper-division physics classes at three very different institutions in the United States, including large public research universities and a small undergraduate-only liberal arts college.  The courses they are enrolled in are common among physics departments, frequently enroll 15-25 students at research institutions and 5-10 students at small liberal arts colleges, and include both theory and laboratory classes.  Our data include in-class and out-of-class observations of students working in small groups and oral exams with individual students which are part of their course grades.  Though some of these courses are taught in a highly reformed manner, others are largely traditional.

The out-of-class observations were conducted as part of optional ``Homework Help Sessions'' wherein students discuss homework problems with classmates and a graduate student teaching assistant (TA).  In a typical HHS, 2-4 students sit around a table with a large table-based whiteboard.  Unlike a typical recitation, the TA does not present problems on a chalkboard, and unlike typical tutorials, there is usually only one group of students so the TA stays with them constantly.  An unattended video camera records the group.  HHS are described more fully in \cite{Sayre2008}.

\subsection{Data selection and analysis}
The ethnographic research methodology originates from the discipline of anthropology\cite{Pirie1997b, Garfinkel1967} but has since been applied to the educational setting in attempts to characterize relationships and events that occur in different educational settings \cite{Collins2004, Brown2004, Hollan2000}. It is a common methodology used when trying to understand community life \cite{Marcus1998, Barab2002} and is generally concerned with the sociocultural features of an environment, how people interact and their discursive practices \cite{Brown2004, Pirie1997b} or essentially investigate the ``culture of the classroom''. Ethnography can be used to characterize various relationships and events\cite{Collins2004, Brown2004}.  

Ethnography typically draws its data from a number of sources in order to get a more complete picture of the culture of the classroom but also to attempt to overcome some of the weaknesses of subjectivity through triangulating multiple viewpoints \cite{Ernest1997}. For the larger project, data were drawn from several sources, including video-recorded observations of students in classrooms and class-related settings, interviews with individual students and small groups, and artifacts of their written work. 
Because the focus of this paper is speech in natural settings, we focus on the observational data here.

In the course of our broader study, we noticed that students frequently spontaneously reflect on their learning and what they're supposed to be doing, and that their reflections carried elements of the physicist speech genre.  Intrigued, we started an investigation of how students' BESM talk indicated appropriation or resistance to TLP.  Over several months, three investigators watched archival video of students interacting in Advanced Laboratory classrooms,\cite{Irving2104} working in homework help sessions,\cite{Sayre2008} and performing physics in oral exams.\cite{Sayre2014Orals} The video was not acquired for this purpose; however, because it represents a wide range of natural settings for physics students, we believe that it is well-suited and appropriate to our task of investigating students' speech ``in the wild''.

Because BESM talk is definitionally talk, we looked for episodes when students were speaking a lot (as opposed to when they were principally writing in their notebooks\cite{Wolf2014JustMath} or working with equipment).  Because BESM talk is definitionally brief, we further refined our search to episodes in which students' speech was liberally sprinkled with BESM talk. Additionally, because classroom video is often noisy -- many students, lots of cross talk or equipment noise, other things in frame or critical objects out-of-frame -- we selected episodes with clear sound and all critical objects in frame.  These three selection criteria occurred simultaneously.

We closely examined video of students' discourse\cite{Gee2005} -- their words, gestures, tone, prosody, etc -- and interactions to understand how they created and communicate meaning\cite{Parnafes2013, Berland2012, Scherr2009} around BESM talk.  
Through intensive discussion of themes\cite{Jordan1995}, two kinds of BESM talk emerged (Section \ref{sec:besmkinds}). In generating our categories, we sought a tractable number that would describe the variation in students' BESM talk while at the same time functionally help us sort utterances into ``kinds'' roughly evenly. Three additional researchers (six total researchers) came to consensus about the kinds of BESM talk and the instances thereof through careful interrogation of the episodes presented below (Section \ref{sec:examples}).  We chose these episodes for deeper analysis because they represent a wide range of settings and illustrate the different categories of BESM talk; our goal here is not to show the prevalence of BESM talk, but to introduce a meso-theoretical construct for analysis of student thinking.  We connect students' BESM talk to how they were TLP through both critical examination of the literature on TLP and our own cultural sense of what it means to be TLP.

\section{Kinds of BESM talk}
\label{sec:besmkinds}

We identify two basic kinds of BESM talk within the physicist speech genre: ``expectations BESM talk'', and ``self-efficacy BESM talk''.  Table \ref{tab:BESMkinds} contains the different types of BESM talk.  

These BESM talk categories, taken as a whole, overlap with the categories drawn from beliefs inventories like the CLASS and MPEX.  While the CLASS categories are drawn from factor analysis of CLASS responses and MPEX categories are drawn from \textit{a priori} decisions about how MPEX statements should be grouped, our categories emerged from student utterances in natural settings.  These different methods naturally produce different basis sets for epistemological commitments; however, we believe that they are all looking at the same kind of thing and have similar span.

Expectations BESM talk generally encompasses CLASS and MPEX categories about physics in the real world, connections among principles, sense making, and problem solving.  Self-efficacy BESM talk relates to CLASS and MPEX categories about effort and aptitude.  Because individual statements on the CLASS correspond to multiple CLASS categories, it is impossible to exclusively assign each CLASS statement to a BESM category.

\begin{table*}[hpbt]
\caption{Kinds of BESM talk}
\begin{center}
\begin{tabular}{|ll | p{5cm} |}
\hline
Category&Subcategory&Examples\\
\hline
\multicolumn{2}{|l|}{Expectations: students' expectations about technical and cultural physics practices}&\\
&Understanding&
  ``I get it\dots"\newline
  ``That makes sense\dots"\newline
  ``There we go"\\ \cline{2-3}
&Confusion&
  ``I don't understand\dots"\newline
  ``This doesn't make sense\dots"\newline
  ``I've gotten stuck on\dots"\newline
  ``I guess\dots" (in some cases)\\ \cline{2-3}

&Spotting Inconsistencies&
   ``I've done something silly\dots"\newline
  ``I missed a negative sign here\dots"\\ \cline{2-3}

&Generalized Expectations&
  ``We expect it to be\dots"\newline
  ``It should be\dots"\newline
  ``As it should be\dots"\\ 

\hline 
\multicolumn{2}{|l|}{Self-Efficacy: students' belief in their own abilities}&\\
&&
  ``I don't know how to\dots"\newline
  ``If we knew\dots"\newline
  ``I am not good at\dots"
\\
\hline

\end{tabular}
\end{center}
\label{tab:BESMkinds}
\end{table*}%

\subsection{Expectations BESM talk}
Expectations BESM talk encompasses students' expectations about the technical and cultural practices of physics (``physics expectations'').  Because these expectations are often implicit and may change depending on the situation, we are especially interested in their physics expectations while they actually do physics in natural settings such as physics classrooms and laboratories, or while working on their homework. 

Prior work on student expectations in physics has focused on a generalized set of expectations about the nature of physics as a field (e.g. ``physics is hard and doesn't matter outside of physics classes''), or about students' generalized expectations for their own behavior in physics classes (e.g. ``I need to memorize a lot of things for this class'')\cite{Omasits2005, Elby2001, Redish1998, zwickl:442, Hazari2010a}.  Another strand of research concerns students' in-the-moment epistemological framings\cite{Kaczynski2013, Gupta2010, Hutchison2009, Louca2010a}, including both their expectations about physical systems (e.g. ``the block should slow down'') and their framing of instructional activities (e.g. ``we're supposed to be doing the worksheet right now'').  Following a restricted interpretation of this latter strand, we focus on students' expectations about physical systems as they show up in the moment.  

We identify four subcategories of physics expectations:
\begin{description} 
\item[Understanding] is when students make it known that they feel like they understand something in a problem. This typically comes as students follow through a problem and often times is coupled with cues of future behavior. Students make comments along the lines of ``\besm{that makes sense}'' and occasionally explain why it makes sense. This follows along the same lines as their expectations in physics. Unlike future behavior, understanding usually follows sense-making.
\item[Confusion] is when students comment that they are not understanding the material used to solve a problem or a concept from the course. These confusions are important, especially when students can identify why they are being confused, whether it's something mathematical they do not understand, something when transitioning from math to physics, or something that is a social convention within physics that they don't understand. For example, students might say ``\besm{I don't understand why we did this}'' or ``\besm{I think I've confused myself}''.  
\item[Spotting (In)consistencies] is when students spot inconsistencies within their work, or comment on expected consistencies in it. Inconsistencies can sometimes come with a statement of confusion but is often separate from them. A cultural practice among physicists is to often check the reality of their answers to check the consistency of their thinking as they are doing physics. Inconsistencies can often arise due to a mathematical error, unnecessary assumptions, or a mistake in the actual physics. All of these can lead to answers that are nonsensical physically, such as particles traveling twice the speed of light, or are nonsensical mathematically, such as $7=3$. 
\item[Generalized Expectations] is when students talk about their expectations for the behavior of physical or mathematical systems, or of characteristics of physics as a technical endeavor in general.  For example, students may say that ``\besm{we expect} that far away from a charge distribution the electric field goes like $1/r^2$.''  They may also make more general statements of expectations: students may say that ``higher order terms in a series expansion \besm{should} get smaller in general'', reflecting the physics norm that series converge. 
Mathematically, there's no requirement for a series to converge and mathematicians do not generally expect that behavior; the expectation that series should converge and therefore this one will as well is a generalized expectation.   
\end{description}

We realize that these descriptions admit some ambiguity; the following sections discuss some ambiguous cases (\ref{sec:ambig}) and the remainder of the paper is devoted to illustrating examples of BESM talk and its relation to TLP.

\subsection{Self-efficacy BESM talk}

Self-efficacy BESM talk positions the student to or within the field of physics. 

In self-efficacy BESM talk, students talk about their belief in their own abilities or reflecting on their efficacy. This often comes in the form of a statement about one's ability to do a task or a statement about their ability (``\besm{I am really good at math}'' or  ``\besm{I am just not good at doing experiments}''). These statements are important as they often give insight into students' affect to particular aspects of doing physics or how they position themselves to the subject overall.   Compared to confusion BESM talk or understanding BESM talk, these phrases are more generalized statements about their abilities.  Though they are co-located with specific physics content, their scope is larger.

\subsection{Ambiguous cases}
\label{sec:ambig}
\subsubsection{``I don't know''}

There are some phrases in English, such as ``I don't know'' or ``I guess'', which appear on first blush to be metacognitive in nature.  In their usage, they are sometimes genuinely metacognitive -- they reflect an actual state of not-knowing -- such as ``I don't know how to solve this problem'' or ``I don't know who he is''.  However, sometimes they are used as weasel words to indicate disagreement with previous statements or speakers.  For example, ``I don't know if it's a good idea to drive this fast on a winding country lane'' really means ``It is a bad idea to drive this fast on a winding country lane''.  

In determining the uses of these phrases, we examined the surrounding context, tone, and prosody of the students who uttered them, as well as the reactions of their peers in the conversation (if available).  However, because the phrases are ambiguous, there are times when each of us felt like we couldn't tell if they indicated the speaker didn't know, and after discussion, we still couldn't tell.  For those cases, we chose a conservative approach and did not count the phrase as BESM talk. We note that this ambiguity may be deliberate on the part of the students as a face-saving measure, and our conservative approach probably under-samples the actual occurrence of BESM talk in these environments.

\subsubsection{Generalized expectations vs. making hypotheses}

A typical behavior for physicists when performing experiments is make a hypothesis or prediction about the behavior of the system.  Making a hypothesis and expressing a generalized expectation are two different (but related) activities. We want to separate the act of making specific predictions from the act of verbalizing generalized expectations about the nature of physical systems and physics formalism.  

Predicting that ``this ball will fall" is not metacognitive; it is a statement about future events which does not position the speaker as an actor in the system.  Additionally, it's not a generalized expectation because it is confined to this system and this ball.  In contrast, saying ``we expect higher order terms to die off'' is metacognitive: the speaker literally talks about her own thinking (``we expect"). It's also generalized expectation, not limited to the specific series in question, about how series approximations usually work in physics contexts.  

The distinctions between expressing generalized expectations and making hypotheses can be difficult at times.  As in the case of ``I don't know'', researchers must rely on the context and tone of students' utterances to help determine whether they are BESM talk or something else.  Some researchers may decide to be conservative and others more liberal; we contend that the decision to look for and value students' BESM talk -- no matter how sparse it may be -- is valuable for both researchers and instructors who are interested in whether and how students TLP.

\section{Examples of BESM talk}
\label{sec:examples}

In the remainder of this paper, we present several episodes of students engaging in BESM talk while doing physics in several different physics learning environments.  The students exhibit different levels of appropriation of TLP through their BESM talk.

In the first episode, ``Bill" works in a HHS to understand the signs of the forces in one dimensional damped harmonic motion.  In the second, ``Zeke" works in an oral exam to understand the signs of the forces for a block attached to two springs on a rotating platform. In the third, ``Bob" works on a electromagnetic experiment and reflects on his own shortcomings when it comes to this particular type of experimental work. Both Bill and Zeke were selected for analysis because they are unusually verbal in their discussions compared to their peers. In contrast, Bob is an example of a student who is not particularly verbal but still engages in spontaneous metacognitive talk. All students are fluent English speakers.\footnote{Of all the data in our video library, almost all students are fluent English speakers; there aren't enough non-fluent speakers with clear audio to investigate how fluency affects BESM use.} The episodes selected for this paper were chosen as rich cases of BESM talk from natural settings in physics classes.  

A typographic note: when BESM talk occurs within larger blocks of transcript, we denote it \besm{like this}.  Within a block of transcript, dash denotes interruption, comma denotes a small pause, period denotes a longer pause, and ellipses a long pause.  When we have removed words or sentences of transcript for brevity in reporting, we denote their removal with ellipses in square brackets: [\dots]; changed or added words are also set off by square brackets, while gestural information is set off in parentheses.

The technical details of the problems presented in these examples are both deeply important to their solutions -- their solutions hinge on coordinating technical and cultural aspects of the problems -- and beside the point of this paper.  Where necessary, we have presented a brief overview of the physics involved in their problems.   

\subsection{Zeke}
\label{sec:zeke}
\begin{figure}[tb]
\begin{center}
\includegraphics[width=7cm]{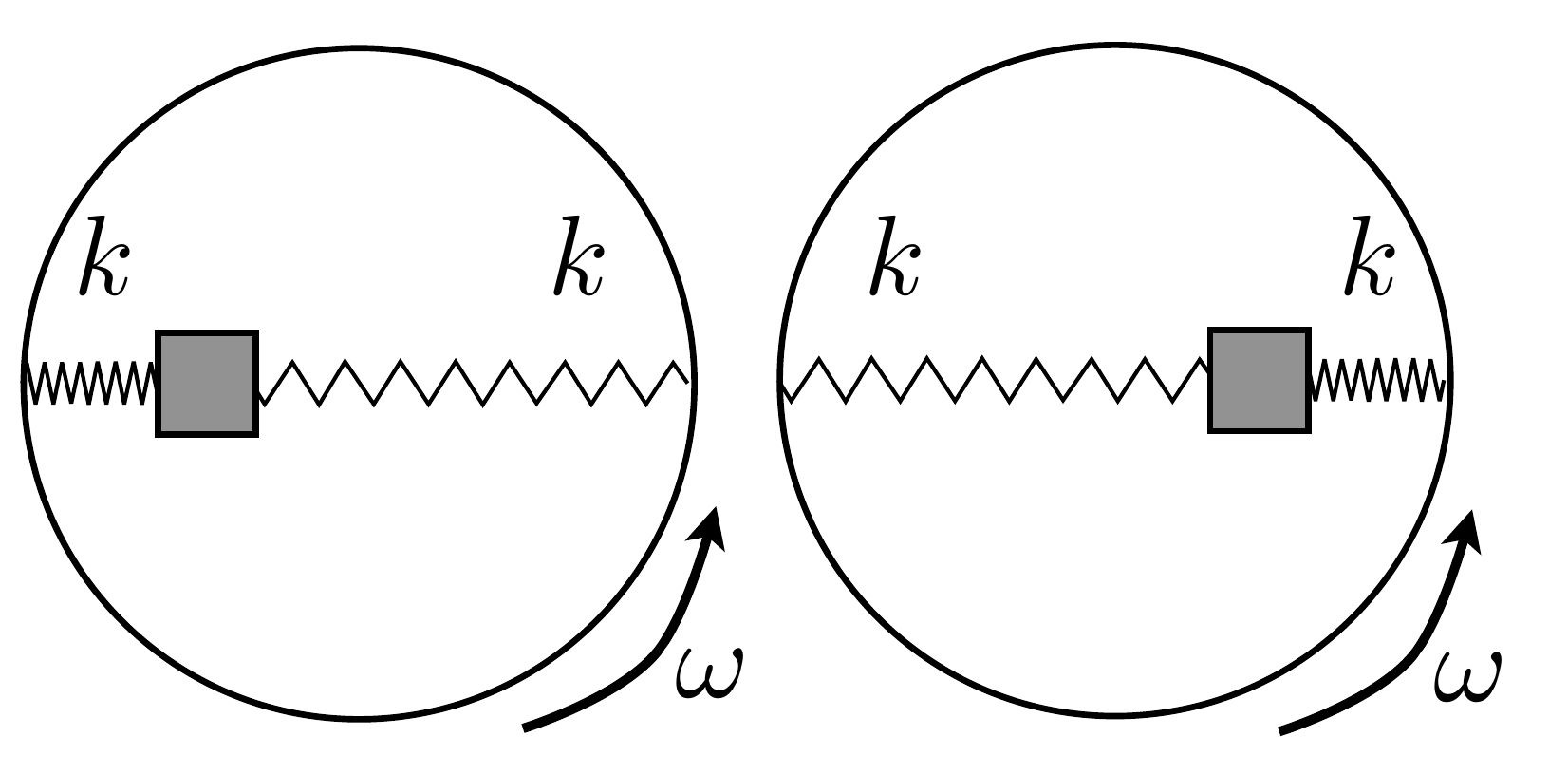}
\caption{Left: the initial diagram.  A block is attached to two springs and displaced to the left.  Each spring has spring constant (stiffness) $k$, and the circular rotating platform spins in the plane of the page at angular speed $\omega$.  Right: Zeke's revised diagram, in which he moves the block to the right of equilibrium.}
\label{fig:Zeke-springs}
\end{center}
\end{figure}

``Zeke'' is enrolled in a similar Classical Mechanics course to Bill (below) and Julie, but at a small liberal arts college.  Zeke is unusually verbal and forthcoming about his ideas.  He is easily the top of his class and especially competent mathematically.  This episode is taken from an oral exam in the second month of the course.  There are two participants, Zeke and the examiner, and the episode is largely a monologue by Zeke.  Zeke is working on a problem where a block is attached to two springs and constrained to move in one dimension on a horizontal rotating platform, as shown in Figure \ref{fig:Zeke-springs}.  

Zeke spends about 17 minutes solving the whole problem. Initially, the problem is drawn with the mass to the left of equilibrium, but Zeke quickly moves it to the right of equilibrium. He considers a block attached to one spring on a non-rotating platform.  Then he adds the second spring and checks for consistency between his sketch of the physical scenario, his free-body diagram, his algebraic signs of forces, and his embodied sense of how the forces should push and pull.  He allows the platform to rotate, and solves for the conditions under which the block undergoes harmonic motion.  This chain of reasoning -- breaking the problem into these specific simpler ones and adding new features -- is very physicist-like.

\begin{table}[hpbt]
\caption{Zeke's BESM talk in a 3.5-minute episode.}
\begin{center}
\begin{tabular}{|l|l|}
\hline
Category&Count\\
\hline
Generalized Expectations&4\\
Understanding&6\\
Confusion&2\\
Spotting (in)consistencies&3\\
\hline
\end{tabular}
\end{center}
\label{tab:zekebesms}
\end{table}%
%

In this paper, we focus on 3.5 minutes near the start of the problem where Zeke considers the signs of the forces for each spring in the two-spring non-rotating system.  Over the course of these 3.5 minutes, we identify 15 instances of BESM talk, as summarized in Table \ref{tab:zekebesms}.  Zeke's talk is rife with examples of expectations BESM talk, though the subtleties of his TLP bear close scrutiny.

Zeke has expectations of consistency between algebraic expressions and the physical systems they represent, a crucial part of thinking like a physicist: 
\begin{description}
\item[Zeke:] The force on this [left] spring.  Um.  I'm going to write something down but \besm{I don't think it's right}.  \besm{I'll have to think about it}.  Negative [writing] $k$. $x$.  So ok, so if $x$ is negative, then \besm{we expect} $k$ to push\dots um, ok.  So in this situation\dots  
\end{description}
However, at this point he gets stuck. Zeke finds it difficult to work with the mass on the ``negative'' side of equilibrium.  He expects that his representations should be consistent, but the sketch seems backwards to him. He moves the mass to the right, saying that he ``just [wants] to make the sign \besm{make more sense}''.  

Zeke's rigid, implicit coordinate system where positive is the rightwards direction is typical of introductory and intermediate mechanics students\cite{Sayre2008}.  An implicit coordinate system is typical of physicists, and we interpret his (lack) of speech in making it explicit as part of the physicist speech genre. Zeke's BESM talk here indicates that he thinks about coordinate systems similarly to his peers at the intermediate level, indicating appropriate mastery of physical technical content.  Later in the problem, he fluidly shifts to a polar coordinate system to deal with the rotating platform.  This shift in coordinate systems, using whichever system is easiest for the (sub)problem at hand is typical of advanced physics students and physicists.  


Immediately after fixing the position of the mass on the picture, Zeke considers the force by the (now extended) spring on the left, followed by the force by the (now compressed) spring on the left.  
\begin{description}
\item[Zeke:] So this, this uh [left] spring is going to pull this way [to left]. We have a positive $k$.  So negative $kx$ \besm{makes sense} because we have a positive $k$ displacement and um it's going to be pulling it backwards in the negative direction.  This one [the spring on the right], it's going to be pushing, so it's going to be pushing at\dots um\dots (mumbling: $kx$ is this right).  Um.  As $x$ gets bigger [points right], this, the force gets bigger, by a factor of $kx$.  So it's going to push this way at $kx$, \besm{that's correct}. 
\end{description}
In these two segments, we notice that Zeke's problem solving approach is very physicist-like\cite{VanHeuvelen1991}. He draws a diagram, he breaks up the problem into smaller ones, he tries to first explain physically what happens, etc.  He overtly connects these ways of knowing to his mathematical ways of knowing, and seeks consistency between representations.  Using an Airey and Linder perspective, he has achieved discoursive fluency through use of many facets of a disciplinary way of knowing.\cite{Airey2009}  His use of BESM talk makes evident his expectations, and allows us to judge whether he is merely aping the behavior of a physicist or if he has appropriated these ways of knowing.



Zeke continues:
\begin{description}
\item[Zeke:] Ok, so let's see \besm{if this makes sense}.  So if [the mass] $m$ goes this way [to the right], positive $x$ displacement.  \besm{We expect,} we have a positive $x$.  This [algebraic sign of force from left spring] is negative, points that way [leftwards], \besm{that makes sense}.  [For the other spring,] we've got a positive $x$, this [algebraic sign of force from right spring] is negative, points that way [left], \besm{makes sense}.
\end{description}
At this point, Zeke pauses, considers his work, and steps away from the board.  Still facing it, he says ``\besm{No, I've done something silly.}''  We interpret this BESM talk as spotting an inconsistency in his reasoning. It is interesting to note that Zeke's reasoning -- though difficult to read -- is both complete and largely correct at this point.  There are no inconsistencies between his sketch, his algebra, his verbal reasoning, his embodied pointing, and physicists' conventions about these representations.  

\begin{figure}[tb]
\begin{center}
\includegraphics[width=4cm]{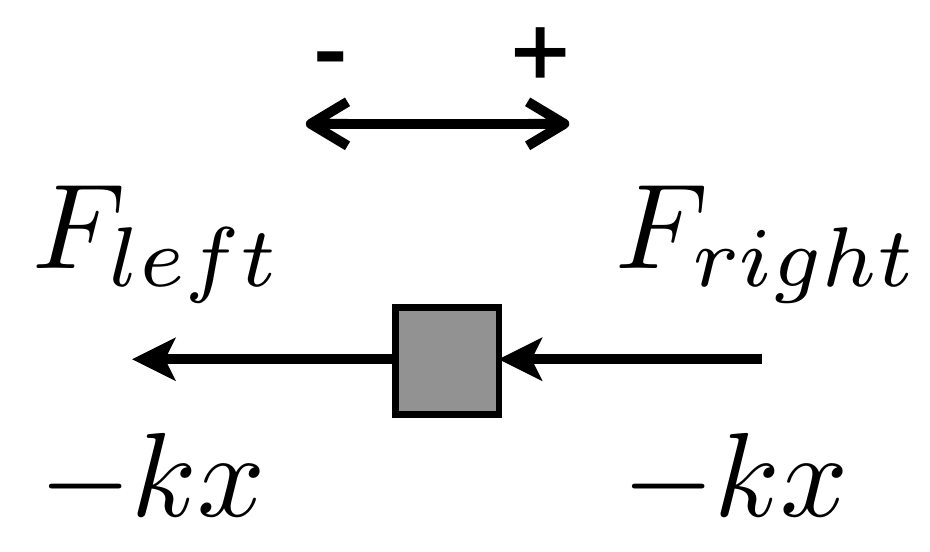}
\caption{Zeke's free-body diagram.}
\label{fig:Zeke-FBD}
\end{center}
\end{figure}

His free-body diagram has two arrows pointing to the left to represent both leftwards forces (as it should); however, one of the arrows comes out of the center of the diagram and the other points into the center (as in Figure \ref{fig:Zeke-FBD}), contra to the convention that all arrows should point out of the center.

This small inconsistency is enough to make Zeke erase the negative sign associated with the algebraic representation of the force on the right.  He continues checking for consistency and reflects on his own confusion:
\begin{description}
\item[Zeke:] No negative there because\dots at positive $x$ displacement\dots  No.  \besm{I was right}\dots It was, it was this [free-body diagram] that \besm{was making it confusing.}  This [an arrow on the free-body diagram] should be pointing this way [leftwards and towards the box].
\end{description}
Having resolved the inconsistency, Zeke checks again.  His BESM talk indicates that he is seeking inconsistencies and that he has located not just an inconsistency, but a cause for his confusion.  He fixes the errant arrow on his free-body diagram.

At this point, the series of checks between representations has become a stylized litany.  His gestures have become smaller and less precise, indicating that he is no longer using them as mindful sense-making tools.  Each phrase is said rapidly, and his tone has begun to become more even, as if he is checking off items in a mental checklist:  
\begin{description}
\item[Zeke:] Alright, so: positive $x$ displacement [points right], negative $kx$, points that way [points left with other hand].  \besm{That's what we want.}  [For the other spring] Positive $x$ displacement \dots [erases a negative sign for the right force] \dots (mumbling: positive $x$ displacement, negative $kx$ points that way)  positive $x$ displacement, positive $kx$. No, because \besm{we want it} to push back. [replaces negative sign]. 
\end{description}
In this segment, Zeke repeatedly uses ``we want'' to indicate his expectation that these representations should be consistent with each other and with his kinesthetic sense of which way the forces should push and pull. He once again considers making the sign of algebraic formulation of the force on the right positive, and rejects it, this time referencing his expectation that the force should be pushing the mass backwards towards equilibrium, in the negative direction.  Zeke's expectations BESM talk has changed from seeking inconsistencies in his reasoning to verifying that his reasoning is consistent.  It is interesting to note that his final, convincing piece of reasoning (``we want it to push back'') draws on his kinesthetic sense of pushes and pulls rather than memorized conventions about algebraic signs.  We interpret this piece to mean that Zeke's physical intuition may drive his mathematical understanding, a sophisticated relationship that indicates he is thinking like a physicist.

At the end of Zeke's error-checking, he turns around, further implying that he was working for himself and not the benefit of the examiner, and says ``Ok.  I'm happy now.'' He and the examiner exchange some cursory words to establish that Zeke has written $-kx$ for each force and that each $x$ or $k$ is the same as its twin.  

Zeke's body language when he turns around is in contrast to his body language during his error-checking.  Throughout his error-checking, Zeke continues to face the board, away from the examiner. His gestures are largely obscured from the examiner, and his body blocks most of his writing on the board.  His body language indicates  that he is working through this problem to satisfy himself.  This claim is supported by his speech (note the contrasts to Bill's speech, below).  Zeke does not address the examiner nor ask her if his reasoning is correct.  The examiner, meanwhile, does not speak at all unless she feels Zeke is addressing her, and she does not speak during his extended error-checking. 

Zeke's repeated use of ``makes sense'' and his repeated, insistent checks that his representations -- algebraic, diagrammatic, pictoral, and embodied -- are consistent indicate very strong appropriation of the physicists' expectation for consistency across representations.  

\subsection{Bill}
\label{sec:bill}
``Bill'' is enrolled in a Classical Mechanics class at a large research university.  Bill is quite verbal about his ideas, like Zeke, but he struggles a lot more with coordinating mathematics and physics ideas.  This episode is taken from the second month of the course.
Bill and another student work in a HHS with a TA to solve the equation of motion for a damped harmonic oscillator, $F = -cv -kx = ma$.  Like Zeke, Bill divides the problem in to cases and seeks consistency between them; unlike Zeke, Bill needs a lot more guidance from his interlocutors to organize his thoughts and resolve his confusion. 

\begin{figure}[tb]
\begin{center}
\includegraphics[width=7cm]{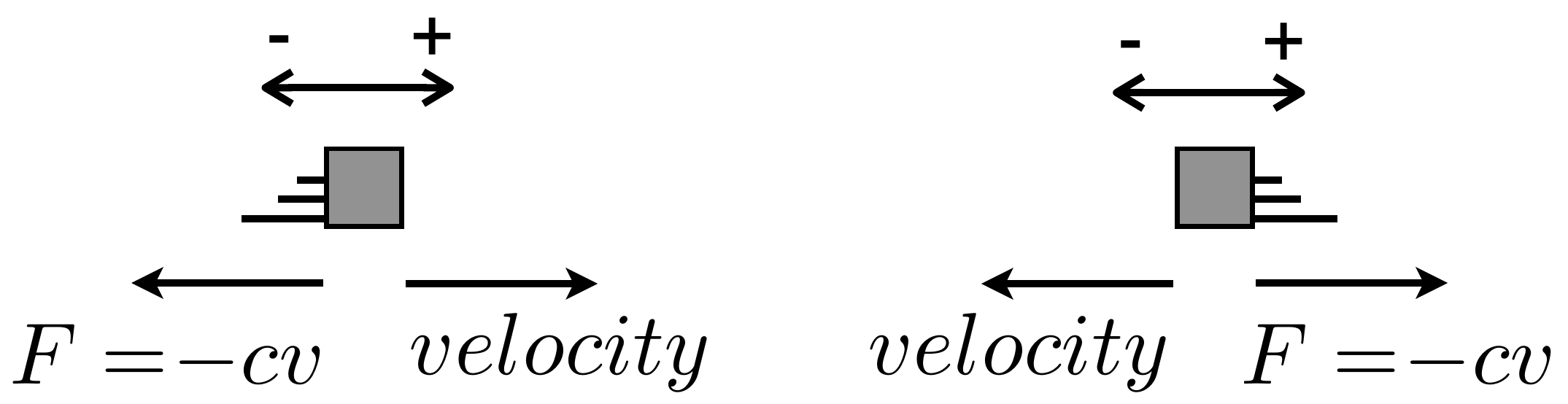}
\caption{Drag forces on a moving block.}
\label{fig:bill-air}
\end{center}
\end{figure}

\begin{figure}[tb]
\begin{center}
\includegraphics[width=7cm]{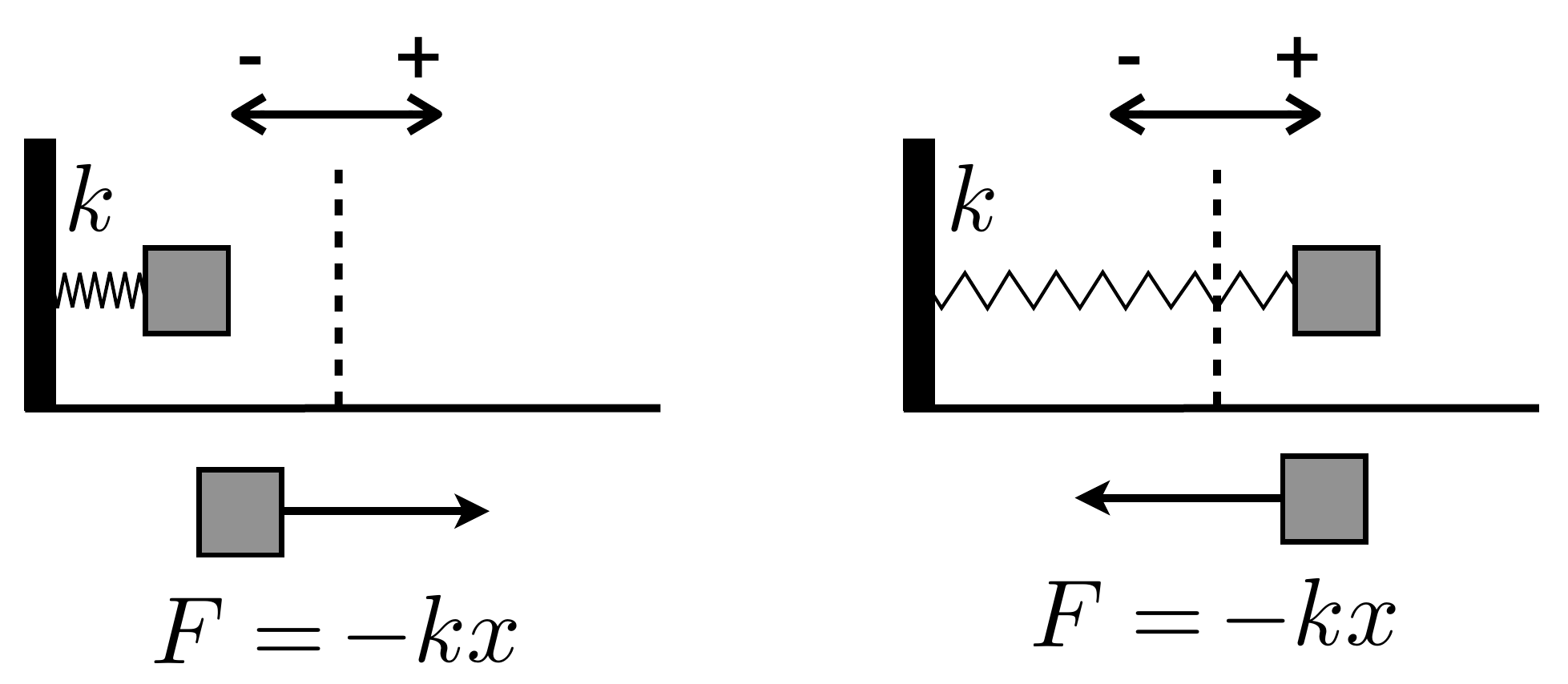}
\caption{Spring forces on a moving block.}
\label{fig:bill-spring}
\end{center}
\end{figure}

The problem that Bill works on is somewhat different than the problem Zeke works on.  Figure \ref{fig:bill-air} shows that when a block is moving to the right, the drag force is to the left.  If we impose a coordinate system where positive is to the right (as Bill does, and as is common for physics students), the velocity $v$ is positive and the force $F$ is negative, so the algebraic sign of the expression for the force, $F=-cv$ must be negative.  If the block is moving to the left, the signs of both the velocity and the force reverse and the algebraic sign of the expression for the force remains the same.  Figure \ref{fig:bill-spring} shows that for the spring force, the direction of the velocity does not matter; instead, the position of the block matters.  When the block is to the left of equilibrium, it compresses the spring.  The spring then pushes it back towards equilibrium, the opposite direction of its displacement.  When the block is to the right of equilibrium, it stretches the spring, which then pulls it back towards equilibrium.  In both cases, the direction of the force opposes the displacement, and the spring force is expressed algebraically as $F=-kx$.

In the episode we analyze here,\footnote{Portions of this episode previously appeared in \cite{Sayre2006}.} the other student has already written down the equation and begun to solve it on a table-based whiteboard in front of all three participants.  Throughout the interaction, most of Bill's speech is technical in nature.  In the approximately six-minute episode with 45 turns-at-talk among the three participants, we count 14 instances of Bill's BESM talk (Table \ref{tab:billBESMs}); almost all the speech in the interaction is Bill's, and the other two participants' turns are generally short.

\begin{table}[hpbt]
\caption{Bill's BESM talk in a six-minute episode.}
\begin{center}
%
\begin{tabular}{|l|l|}
\hline
Category&Count\\
\hline
Generalized Expectations&6\\
Understanding&3\\
Confusion&4\\
Spotting (in)consistencies&1\\
\hline
\end{tabular}
\end{center}
\label{tab:billBESMs}
\end{table}%

Bill's BESM talk tends to cluster within turns-at-talk.  Near the beginning, his statements express both generalized expectations about forces and motion, and confusion about how to solve this problem in particular.  In the middle, his BESM talk indicates that he is still confused, but he's starting to use his expectations of consistency.  By the end, his confusion is over and he has found the consistency he seeks.  

At the start of the episode,
Bill is confused and asks we can ``go over" why the drag and spring forces ($-cv$ and $-kx$, respectively) are negative.   He is ``\besm{a bit concerned}'' about why ``both of these are being negative all the time''. The fact that Bill initiates the interaction indicates that Bill is aware of his lack of understanding, and engages to improve it actively.  

After Bill's initial request, the TA indicates that Bill should start reasoning about the two forces, so Bill starts to explain, and trips himself up. 
\begin{description}
\item[TA (0:20):] Well, tell me what you get so far.
\item[Bill (0:22):] Well, \besm{I know that when} the force is like if you have the spring at [\dots]  so it seems like it is negative on both sides of this, \besm{like its always going to oppose} the motion. [\dots] So it ends up being\dots \besm{I have to start over.}  Acceleration is\dots
\end{description}
In the excised pieces of transcript here -- which is extraordinarily difficult to follow -- Bill talks about the forces to the left and right of equilibrium, or perhaps the forces for leftward and rightward movement.  He confuses everyone: himself, the TA, and our research team.  
Bill's confusion  is evident in his BESM talk, and also in his surrounding language: he repeats ``it seems like'' and ``just'' several times, two phrases which soften his statements.  This piece of BESM talk shows both Bill's awareness of his confusion and his agency to clarify it for himself, two indicators of advanced metacognitive development.  It comes directly after technical content in which he (correctly) reasons about the forces but doesn't directly coordinate his force reasoning with the mathematical statements in front of him.

The TA prompts Bill to look only at the right side of the diagram, and Bill tries again, also confusing himself.  He starts over, explaining that the drag force opposes velocity and the spring force opposes the displacement, a physically sufficient explanation, but one which does not (overtly) include the algebraic signs of the forces.  

Bill reports that, in class, the professor ``did this thing with his thumbs'' (1:27), pointing in different directions and moving his hands from side to side to indicate the forces as a function of position and velocity direction.  Bill is also confused about how this representation helps him understand the signs of the forces, and his BESM talk starting at 1:47 reflects that:
\begin{description}
\item[Bill (1:47):] So, I just, it seems like they can both have the same sign, sign... and \besm{I don't know.}  This one's position and the other one's velocity and...
\item[TA (1:55):] Let's worry about one of them, one of them at a time.  Which one would you like to worry about first?
\item[Bill (1:59):] \besm{I don't know,} just... I'll say velocity (points at -cv).  \besm{I'd just like to prove to myself that its correct for every case} and I don't know, if that's (unclear) 
\end{description}
At the end of his turn at 1:47, Bill trails off.  His tone is dismayed, and his efforts to understand the thumb-pointing representation have not helped him understand the algebraic representation.  However, he still expresses a very physicist-like behavior: splitting the motion into cases, and checking for consistency among those cases.  

\begin{figure}[tb]
\begin{center}
\includegraphics[width=7cm]{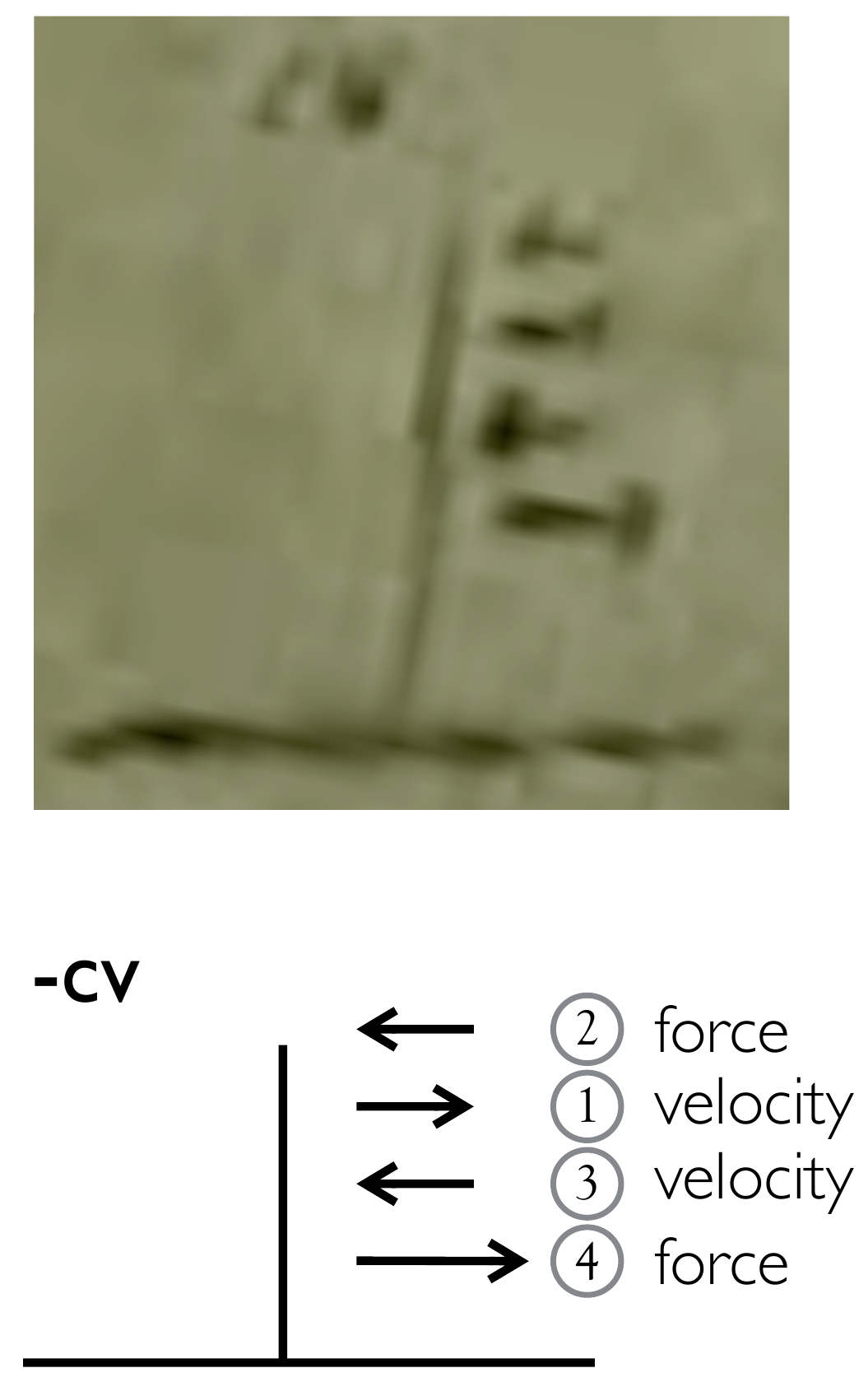}
\caption{Top: Bill's diagram that he draws during this segment.  Bottom: Our sketch of Bill's diagram, shown to compensate for the poor picture quality of the original. Bill names the arrows as he draws them (numbers 1-4 on our sketch): velocity, force, velocity, force.}
\label{fig:bill-diagram}
\end{center}
\end{figure}

To refocus discussion, the TA asks about the drag force first, holding the spring force for later. Bill builds an explanation about the drag force in which he coordinates his reasoning about opposing force and velocity from before with the algebraic sign of the force term by using an implicit coordinate system where positive is rightwards, a box and equilibrium line sketched on the whiteboard, and pointing gestures with his hands to alternately indicate force, velocity, and position. His words do not adequately capture the whole of his explanation, which is more clear in the video.  We present here a lightly-abstracted transcript with gestural information\cite{Scherr2008}; his sketch is shown in Figure \ref{fig:bill-diagram}.

\begin{description}
\item[Bill (2:27):] [when the box is to the right of equilibrium and moving rightwards] This velocity is going to be this way (draws arrow 1, pointing right) and the air resistance is going to be that way (draws arrow 2, pointing left).  Alright, so the velocity is positive and the force is negative (points in opposite directions). And when the velocity becomes negative (draws arrow 3, pointing left), the force is positive (draws arrow 4, pointing right). So it changes the sign around (flips hand over) of this [the v symbol] in here (taps algebraic representation of the force)\dots  Alright?  Is that right?
\end{description}

During his explanation, Bill's attention is on the whiteboard in front of him, his sketching, and his hands. The pauses and connective speech -- "alright, so" and "and when" -- indicate that he is building an explanation on the fly, alternately considering rightwards and leftwards movement and verifying that the algebraic expression works for both. Bill is not performing a previously-derived justification for the TA; instead he is deriving as his speaks.

Bill's explanation is excellent. In it, he displays several physicist conventions which indicate that he is thinking like a physicist.  His implicit coordinate system follows convention: positive is right. He draws a square box to move left and right and a line under it to represent both the coordinate system and the floor on which the box rests.  He indicates the equilibrium position of the box -- even though it is not relevant for this part of the problem -- by drawing a vertical line in the middle of his diagram, simultaneously representing equilibrium and the zero position of his coordinate by defining an origin.  

Additionally, Bill's expectations about the nature of math in physics are physicist-like.  He expects that these different representations should be commensurate.  He considers the two cases separately. Bill engages in the same chain of reasoning that Zeke engaged in, with the same kind of coordinate system and the same kinds of BESM talk.  However, unlike Zeke, Bill does not coordinate as many representations as Zeke does, suggesting somewhat smaller facility with the discourse of physicists.  
 Altogether, Bill's explanation indicates that Bill is taking up both the technical and cultural aspects of becoming a physicist.  

However, despite Bill's earlier BESM talk (before 2:27) which indicates high levels of metacognitive awareness and autonomy, the end of his explanation here is quite different.  We might have expected him to reflect that his explanation ``makes sense" (expectation, understanding subcategory) or perhaps that he is no longer confused and it's ok to continue with the longer problem, much the way Zeke does.  Instead, he asks the TA if his explanation is ``right".  Bill needs the TA to regulate his endpoint and evaluate his reasoning.  

The TA defers Bill's question to the other student, and the three of them have a brief conversation about the effects of reversing Bill's implicit coordinate system.  When they return to the question of why the spring force is negative, Bill builds a similar explanation about the spring force.  His explanation for the spring force displays the same conventions and expectations about the nature of maths in physics.  However, it is much faster, both in speaking speed and number of words, and it uses more pronouns and deictic demonstratives:
\begin{description}
\item[Bill (4:41):] When this is always positive, it's always, it's always positive on [the right] side and the force is always negative, \besm{as it should be}.  \besm{So it should always be} negative since $K$ is positive and $X$ is positive.  And then when it's on [the left] side, the force is always going to be in the positive direction and then this is the (unclear)\dots force\dots Okay.
\end{description}

Bill's BESM talk in the middle of his explanation (``as it should be") indicates his expectations for consistency across representations, and the following BESM talk (``So it should always be") indicates that he expects the sign of the force to be the same on both the left and right of equilibrium. His expectations are explicit here, whereas in his explanation about the drag force they are implicit.  We interpret this change, together with his increased speed and demonstratives, to mean that his expectations are stronger in this case, perhaps because he was successful in this chain of reasoning for the drag force.  

At the end of Bill's explanation, he trails off.  His final ``Okay" is an evaluation of the problem as a whole: his tone says he is satisfied with his answer, and it is permissible for the group to move on.  His verbal description seems incomplete (to both researchers and the TA in the interaction), and the TA expresses that it confused her.  Bill responds with ``No. I get it." in an authoritative and final tone, indicating strongly that he has completed his sensemaking and no longer needs the validation of the TA.   When she presses him further, he ``explain[s] it to [her]", engaging in a performance whose prosody and tone are very different than the sensemaking he engaged with earlier.

Altogether, Bill's BESM talk indicates his growing confidence and autonomy, his expectations about consistency across representations and about the role of mathematics in physics, and his uptake of physics conventions -- all indicators that Bill is learning to think like a physicist.  


\subsection{Bob}
\label{sec:bob}

``Bob'' works with two fellow students in an Advanced Laboratory course.  In the course, each laboratory has a lifetime typically of four sessions, each 3 hours long, spread out over two weeks.  Students work in groups of three, each group on a separate experiment, in the same room together. An instructor circulates in the room.  Typically, each group is closely adjacent to one or two other groups.\cite{Irving2104}

Bob's group is just starting an experiment to measure the charge-to-mass ratio of an electron using varying electric and magnetic fields in an evacuated tube, a method pioneered by Thompson and refined by Hoag in the early 20th century.  The equipment for this experiment is quite fussy, and includes several pieces.  

The students have a somewhat complicated circuit diagram which describes how to connect the equipment, a common tool for physicists to use in setting up experiments.  Bob tells his groups that "\besm{I am not very good at} reading these [circuit] diagrams.''  This piece of self-efficacy BESM talk is a reflection on Bob's ability in reading circuit diagrams.  It allows him to position himself as a non-expert in experimental set-up.  While the rest of his group dives in to connecting equipment, Bob hangs back and crosses his arms.  His gestural signals of non-participation corroborate his BESM talk: Bob doesn't want to set up this equipment because he feels he isn't good at it.

At the time of this episode, the lab group has been working on setting up the equipment for just over a hour without making much progress. The other students in Bob's group and the instructor are together on one side of the experiment set-up table, troubleshooting why some parts of the equipment do not have any power.  Bob observes their actions but does not take part in the equipment set-up process. Bob stands apart from his lab group and near another group, which includes ``Toby''.  Bob and Toby chat about the experimental set-up that Bob's group -- but not Bob -- is working on.

\begin{description}
\item[Toby:] (directed at Bob and pointing to the experiment set-up) There are so many different parts to that experiment.
\item[Bob:] (looks at the set-up and shakes his head up and down in agreement) \besm{I have no idea what is going on}.
\item[Toby:] (laughs)
\item[Bob:] (pauses, looks into the distance) \besm{I'm not really good at it}, [the instructor] set that up.
\end{description}

Bob's last BESM statement relates to his earlier utterance about circuit diagrams, but it is a stronger statement of self-efficacy. We infer that ``it'' in Bob's last statement means setting up experiments or manipulating equipment in general -- a broader scope than ``reading [circuit] diagrams'' -- because the second half of Bob's statement (``[the instructor] set that up'') doesn't make sense if it's about circuit diagrams, but does make sense if it's about equipment.  Additionally, Toby was not part of the earlier conversation about circuit diagrams.
 
Bob seems to be aware of a gap in his ability to set-up experimental apparatus and indicates this through BESM talk not only indicating that he currently does not understand what is going on but that he is also not typically good at setting up experimental equipment. Bob's statement is spontaneous:  Toby does not ask Bob to reflect on his abilities in lab, nor to reflect on his understanding of the current experiment that he is working on.

Bob's self-efficacy BESM talk allows him to position himself as a non-expert, and it helps explain why he excuses himself from setting up the lab equipment.  In a laboratory course like this one, using the equipment and setting up experiments are major components of the course.  Unsurprisingly, because Bob sets himself up as a detached non-expert, he does not participate fully in the course and later drops it.

\section{Discussion}

Julie, Bill, Zeke, and Bob all display BESM talk in their respective physics settings. 

Bill and Zeke use expectations BESM talk to connect their understanding of physical quantities and algebraic expressions for forces.  They are engaged in active problem solving.  They exhibit many physicist-like expectations about consistency between representations.   Bill and Zeke's BESM talk shows them to be thinking like physicists. They have appropriated the cultural tools appropriate to coordinating representations in each of their problems, and their expectations are commensurate with physicists' expectations of physical systems and the formalism used to describe them.  

In contrast, Julie and Bob both use self-efficacy BESM talk to reflect negatively on their role and efficacy within physics. Bob reflects that he is ``not really good at'' experimental set-up, a core competency for experimental physicists and a crucial component of the lab-based class in which he is enrolled.  Bob uses his BESM talk to explain why he does not delve into set-up with the rest of his group.  Julie's BESM talk reflects more broadly on her understanding of physics in general: instead of merely positioning herself as ``not really good at'' an experiment, she ``doesn't know enough physics apparently''.  Julie and Bob use BESM talk to position themselves as outside of physics, and thus we interpret their words as resistance to appropriating physics cultural tools.

It is tempting to think that students who engage in more BESM talk might be thinking more like physicists.  This comparison is difficult to make because some students simply talk more than other students, and some students are more willing to verbalize BESM talk while working together.  It is dangerous to interpret willingness to verbalize as ability to verbalize.  For this reason, students should not be rated as high or low physics thinkers purely on the basis of the \textit{quantity} of their BESM talk.  Nonetheless, BESM talk remains a useful marker for examining students' expectations about the technical and cultural practices of physics, and for examining their self-efficacy as regards the field. 

Because it occurs spontaneously in natural contexts, BESM talk is useful to researchers interested in environments with high ecological validity, especially ones in which participants work together.  We frame learning as a participatory process in communities, so methods for investigating learning \textit{in situ} are vitally important because one cannot meaningfully separate learning from social context.  

Educators attuned to listening for BESM talk may also use it to assess their students' grasp of the hidden curriculum, particularly the physicist speech genre and what it means to think like a physicist.  In courses for upper-level physics majors, these goals are often integral (if unarticulated) parts of the curriculum.  However, we note a growing collection of physics courses for non-majors whose primary emphasis is on understanding the nature of physics and physicist ways of knowing (e.g. \cite{Wittmann2007, Redish2013a}).  For educators in those classes -- and the researchers who study them -- BESM talk is a tool for noticing how students appropriate or resist appropriation to the cultural tools of physics, even as they are not exposed to the canonical topics in a standard curriculum. 


It is an open question as to how students' fluency in English interacts with their use of the physicist speech genre, or with their use of BESM talk.  If learning to talk like a physicist is an important part of becoming a physicist, do students whose first language is not English pick up culturally physicist clues differently?  First generation college students and other students without ready access to STEM professionals in their daily lives may encounter difficulties appropriating the language of science; their self-efficacy BESM talk may reflect these difficulties (as it did for Julie).  

Alternately, researchers who are interested in student development might use changes in the frequency and character of BESM talk over time an indicator of how students become more physicist like.  By tracking individuals along multiple semesters, through several classes and settings, a broader ethnographic study of student development could look at how expectations about physics change over time, or how students position themselves with respect to the field changes with setting and experience. BESM talk may be a productive element of the speech genre to investigate these questions.   

\section{Conclusion}

As teachers, we are interested in helping students learn to TLP because it is a vital component to becoming physicists.  More broadly, we are interested in helping students learn to think like scientists, even the students who are unlikely to become scientists.  As researchers, we are interested in connecting students' behavior in natural settings to their ideas about physics, both technical and cultural.  

While TLP has been elusive for researchers to define, it remains important to practicing physicists and physics professors.  It bridges the technical and cultural aspects of physics. In our roles as researchers and as teachers, BESM talk allows us to infer how students TLP and how they position themselves in relation to the field.  BESM talk occurs spontaneously in wide-ranging instructional contexts and is not limited to the research laboratory, features which increase its relevance to classroom teachers and increase its ecological validity to education researchers.  Connecting student behaviors to TLP allows us to hone our observational skills and build better models of our students and their learning, which may lead to better curricular design. 

\section{Acknowledgments}
The authors are indebted to insightful colleagues and beta readers who offered criticism on this work (in alphabetical order): Ian Beatty, Warren Christensen, Scott Franklin and SMERC, Sissi Li, and Sam McKagan.  Lauren Harris performed some of the classifications of BESM talk, and the KSU PER group participated in inter-rater reliability testing. We are also grateful to the instructors who shared their classes with us.  

This work was partially supported by the KSU Department of Physics and by NSF grant PHY-1157044.  Data collection for the video library which we plundered for our analysis was originally supported by the Lilly Foundation,  NSF grants DUE-0442388 and REC-0633951, and the KSU Department of Physics.  Any opinions, findings, and conclusions or recommendations expressed in this material are those of the authors and do not necessarily reflect the views of the National Science Foundation or other funders.
  
\ecsend{
\bibliographystyle{plain}

} 

\end{document}